\documentclass[aps,prl,twocolumn,superscriptaddress,longbibliography]{revtex4-2}

\usepackage[utf8x]{inputenc}
\usepackage[T1]{fontenc}
\usepackage{graphicx}
\usepackage{bm}
\usepackage{dsfont}
\usepackage{mathtools}
\usepackage{amsmath,amsfonts,amssymb}
\usepackage{comment}
\usepackage{float}
\usepackage{array}
\usepackage{units}
\usepackage{verbatim}
\usepackage{verbdef}
\usepackage{hyperref}
\usepackage{xcolor}
\usepackage{tikz}
\usepackage{placeins}

\begin{document}
\title{Commensurate-incommensurate Mott transition without magnetic field: emergence of nematic Luttinger liquid in XXZ chain}
\author{Julien Fitouchi}
\affiliation{Quantum Nanoscience, Delft University of Technology, Lorentzweg 1, 2628CJ Delft, The Netherlands}
\author{Natalia Chepiga}
\affiliation{Rudolf Peierls Centre for Theoretical Physics, University of Oxford, Clarendon Laboratory, Oxford OX1 3PU, United Kingdom}

\date{\today}

\begin{abstract}
\noindent We investigate the zero magnetization and ground state phase diagram of a spin-1/2 chain with competing ferromagnetic nearest-neighbor and antiferromagnetic next-nearest-neighbor exchange couplings in the strongly interacting regime. Using density matrix renormalization group (DMRG) simulations, we discover two successive commensurate-incommensurate transitions of the non-conformal Pokrovsky-Talapov universality class, occurring even at zero magnetic field. We argue that the first transition marks the proliferation of solitonic discommensurations (kink-antikink pairs) in a critical phase with central charge $c=2$, emerging from a period-4 gapped phase. At the second transition, an incommensurate quadrupolar (or nematic) Luttinger liquid forms out of a gapped phase separation, via the condensation of two-magnon bound pairs. In the strongly interacting limit, the location of the two Pokrovsky-Talapov transitions can be analytically predicted with degenerate perturbation theory. Our results demonstrate that frustration alone is sufficient to drive continuous commensurate-incommensurate transitions of Mott type and stabilize complex objects with incommensurate quasi-long-range order without doping.
\end{abstract}

\maketitle

\noindent Understanding phase transitions driven by quantum fluctuations in strongly correlated systems is a central topic of condensed matter physics~\cite{giam,tsvelik,sachdev}.
Low-dimensional spin systems including chains and ladders provide a rich platform for exploring these phenomena, especially due to their intrinsic connections to spinless fermion models. Such systems have been extensively analyzed using numerical methods including density matrix renormalization group (DMRG)~\cite{white,schollwock}, as well as analytic approaches grounded in field theories in $(1+1)$-dimensions~\cite{giam,pujol,pujol_2,rao_sen} like the non-linear sigma-model and bosonization.\\
\indent Of particular interest is the characterization of continuous Mott-type quantum phase transitions in (1+1)-dimensions -- transitions separating a critical gapless phase from a gapped one~\cite{schulz,kuhner,giam_mott,nersesyan_mott}. A commensurate-incommensurate transition of this type belongs to the Pokrovsky-Talapov~\cite{kasteleyn}\footnote{also known as Kasteleyn transition in the context of quantum dimers} universality class~\cite{PT,bak,ptrg,lifshitz,KT_to_PT} and arises from the condensation of domain walls -- soliton-like excitations, made favorable by an external parameter such as a magnetic field or chemical potential. At the critical point, the solitons proliferate and produce an incommensurate gapless (floating) phase with a continuously varying wave vector.\\
\indent 
Quantum spin-1/2 chains with competing ferromagnetic (FM) nearest-neighbor (NN) and antiferromagnetic (AFM) next-nearest-neighbor (NNN) exchange form a rich class of frustrated quantum systems, known to host vector-chiral order~\cite{kolezhuk,hikihara1999}, incommensurate correlations~\cite{hikihara2001,essler}, and field-induced multipolar Luttinger liquid phases~\cite{multipolar_LL,lauchli, zhitomirsky}. On the experimental side, cuprate chain compounds with FM NN and AFM NNN exchange interactions have shown evidence of incommensurate magnetic order and multipolar correlations. In LiCuVO$_4$, nuclear magnetic resonance (NMR) measurements reveal signatures of a presaturation phase consistent with nematic correlations~\cite{orlova} as well as incommensurate magnetic order~\cite{masuda}. In LiCuSbO$_4$, muon spin rotation and susceptibility measurements indicate a possible quadrupolar nematic regime~\cite{bosiocic}, while neutron scattering detects incommensurate spin correlations~\cite{dutton}. In AgCuVO$_4$, neutron diffraction combined with muon spin rotation identifies collinear amplitude-modulated incommensurate order perpendicular to the chains~\cite{hromov}. Similar features have also been reported in other materials, such as $\beta$-TeVO$_4$, by NMR and specific-heat measurements~\cite{pregelj}.\\
\indent
In this work, we focus on an XXZ spin-1/2 chain that exhibits such frustration. Its Hamiltonian reads
\begin{equation}
\label{H_spin}
H=\sum_{j}\frac{J}{2}\left(S^{+}_jS^{-}_{j+1}+\text{H.c.}\right)+\Delta_1S^z_{j}S^z_{j+1}+\Delta_2S^z_{j}S^z_{j+2},
\end{equation}
where $S_j^{\pm}=S_j^x \pm i S_j^{y}$, $S_j^{x,y,z}$ are the spin-1/2 operators on site $j$ and $J$, $\Delta_1$ and $\Delta_2$ are the nearest-neighbor (NN) $xy$ exchange, NN $z$ exchange and next-nearest-neighbor (NNN) exchange, respectively. For numerical simulations, we set $J=1$ as the energy scale. By means of the Jordan-Wigner transformation~\cite{giam}, the Hamiltonian \eqref{H_spin} can be mapped onto a spinless fermion model, where the conservation of fermion number corresponds to conservation of total magnetization in the spin language. A previous study~\cite{roux} across different magnetization sectors has identified multipolar Luttinger liquids, but their occurrence at zero magnetization has been overlooked. Another study~\cite{KT_to_PT} focused on the case of $n=1/3$ fermionic filling and reported a Mott transition in the Pokrovsky-Talapov (PT) universality class at strong $\Delta_2$, rather than the commonly observed Berezinskii-Kosterlitz-Thouless (BKT) transition~\cite{Kosterlitz_Thouless}. This behavior was attributed to the interplay between the constraint of fixed magnetization and frustration, responsible for the emergence of an incommensurate Luttinger liquid~\cite{KT_to_PT}, the nature of which, however, remains unknown.

In this letter we will demonstrate the existence of analogous transitions in the zero magnetization sector $m=0$ ($n=1/2$ fermionic filling) and shed light on the nature of the incommensurate gapless phases. Surprisingly, these PT transitions rely on weaker conditions: they do not require restricting to a fixed magnetization sector. At zero magnetization, the phase diagram of the system defined by Eq.~\eqref{H_spin} has been established only in the fully antiferromagnetic case $\Delta_1, \Delta_2 > 0$~\cite{mishra}. Here, we extend this study to the ferromagnetic NN exchange $\Delta_1 < 0$. For DMRG simulations, we use system sizes from $N=300$ to $900$, a singular-value cutoff from $10^{-8}$ to $10^{-14}$, a maximum bond dimension from $\chi=1500$ to $4000$, and an energy convergence criterion from $\epsilon=10^{-9}$ to $10^{-12}$, with fixed boundary conditions $S^z_{1}=-1/2$ and $S^z_{N}=1/2$. We discovered a rich phase diagram featuring two commensurate-incommensurate transitions of Pokrovsky-Talapov type which give rise to incommensurate quadrupolar Luttinger liquids that will be the main focus of this letter. Within the strongly interacting regime $|\Delta_1|,\Delta_2\gg J$, a perturbation analysis provides an analytical prediction for the position of both PT lines, consistent with our numerical findings.\\
\indent
Following the terminology of~\cite{multipolar_LL,lauchli}, we refer to a quadrupolar (or nematic) Luttinger liquid (LL) as a phase in which the low-energy gapless excitations consist of bound pairs of magnons. In this regime, single-spin flips are gapped. Such an LL nature is diagnosed by the quasi-long-range order of the two-spin flip correlation function $\langle S^{+}_{i}S^{+}_{i+1}S^{-}_{j}S^{-}_{j+1}\rangle$ (also referred to as nematic correlations~\footnote{We may write $S^{-}_iS^{-}_j = Q^{--}_{ij}=Q^{x^2-y^2}_{ij} - iQ^{xy}_{ij}$ where $Q^{x^2-y^2}_{ij}=S^{x}_iS^{x}_j - S^{y}_iS^{y}_j$ and $Q^{xy}_{ij}=S^{x}_iS^{y}_j-S^{y}_iS^{x}_j$. Such definitions follow from the nematic tensor defined for spin-1/2 as $Q^{\alpha\beta}_{ij}=S^{\alpha}_iS^{\beta}_j+S^{\beta}_iS^{\alpha}_j-\frac{2}{3}\bm{S}_i\cdot\bm{S}_j\delta^{\alpha\beta}$}~\cite{frustrated_mag}) while the single-flip correlation function $\langle S^{+}_{i}S^{-}_{j}\rangle$ is short-ranged. We will also make use of the nematic density $m_{\mathrm{nema}}=\sum_i\langle S^z_iS^z_{i+1}\rangle/N$ to characterise the nature of the critical phases.

\begin{figure}
    \centering
    \includegraphics[width=\columnwidth]{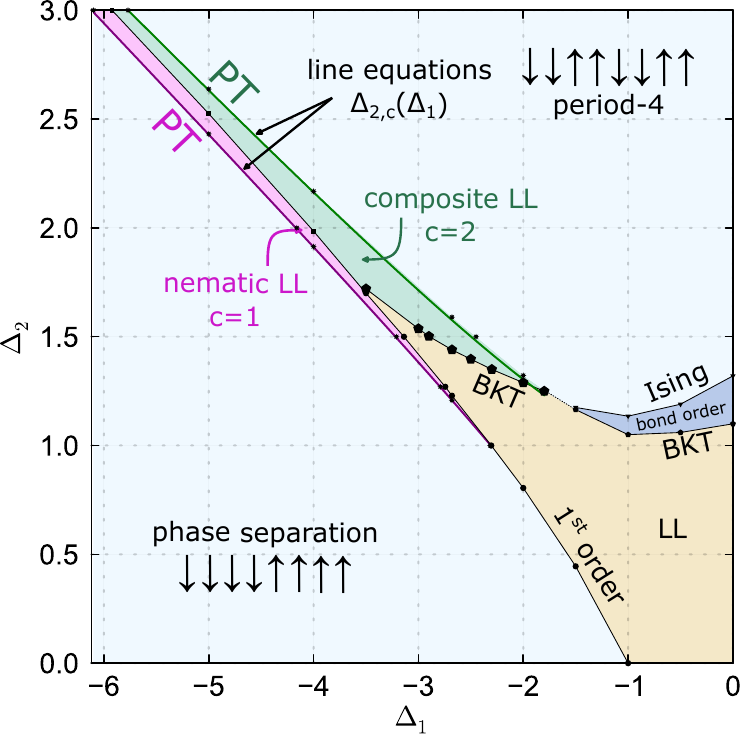}
    \caption{Phase diagram of the system defined in Eq.~\eqref{H_spin} as a function of the nearest-neighbor ferromagnetic exchange $\Delta_1$ and the next-nearest-neighbor antiferromagnetic exchange $\Delta_2$. Lifting the magnetization constraint $m=0$ maps the gapped phase separation onto a ferromagnetic phase and leaves all transition lines unchanged. Three gapped phases -- phase separation, period-4 and bond ordered period-2 phases -- are shown in blue. The latter two are separated by the Ising transition. The first order transition between phase separation and standard Luttinger liquid (LL, beige) is characterized by Luttinger coefficient $K \to \infty$ and velocity $u \to 0$. The transition between LL and bond order is of Berezinskii-Kosterlitz-Thouless (BKT) type with $K_c = 1/2$. In the strongly interacting regime $-\Delta_1,\Delta_2\gg 1$ two Pokrovsky-Talapov (PT) transitions confine a critical sector that hosts a nematic LL with central charge $c=1$ (pink) and a critical phase with $c=2$ (green) which we interpret as a two-flavor LL. Analytic prediction from perturbation theory for the two PT lines are shown in pink and green, respectively. The latter is connected to the standard LL via a BKT transition. The dotted line marks the termination region of the four neighboring critical lines; the underlying scenario remains to be clarified~\cite{XXZ_inprep}. The error bars on the transition points locations are smaller than the corresponding marker sizes.}
    \label{PD_OH_v7}
\end{figure}

\paragraph{Phase diagram.}
Fig.~\ref{PD_OH_v7} provides an overview of the phase diagram for $\Delta_1<0$, $\Delta_2>0$ at zero magnetization. It contains three gapped phases: a phase separation, connected to a standard commensurate Luttinger liquid (LL, beige) via a first-order phase transition; and a pair of ordered phases -- period-4 phase (light blue) and period-2 bond-ordered one (blue) -- separated by the Ising transition. Deeper in the strongly interacting regime ($\Delta_1\lesssim -3.5$, $\Delta_2\gtrsim 1.5$), two Pokrovsky-Talapov (PT) transitions bound a critical region composed of an incommensurate nematic LL (pink) and a critical phase with central charge $c=2$, consistent with the coexistence of incommensurate and single-magnon-type modes (green). Analytic prediction for the two PT lines are shown in pink and green, respectively. The incommensurate mode of the $c=2$ phase is condensed via a Berezinskii-Kosterlitz-Thouless (BKT) transition starting from the standard LL phase. Below, we give numerical and analytical evidences for the two PT transitions and characterize the nematic and composite $c=2$ critical phases. The question of the transitions between the three gapless regimes will also be addressed~\cite{SM}. The rest of the phase diagram including a potential multicritical point will be reported elsewhere~\cite{XXZ_inprep}.

\begin{figure}
    \centering
    \includegraphics[width=\columnwidth]{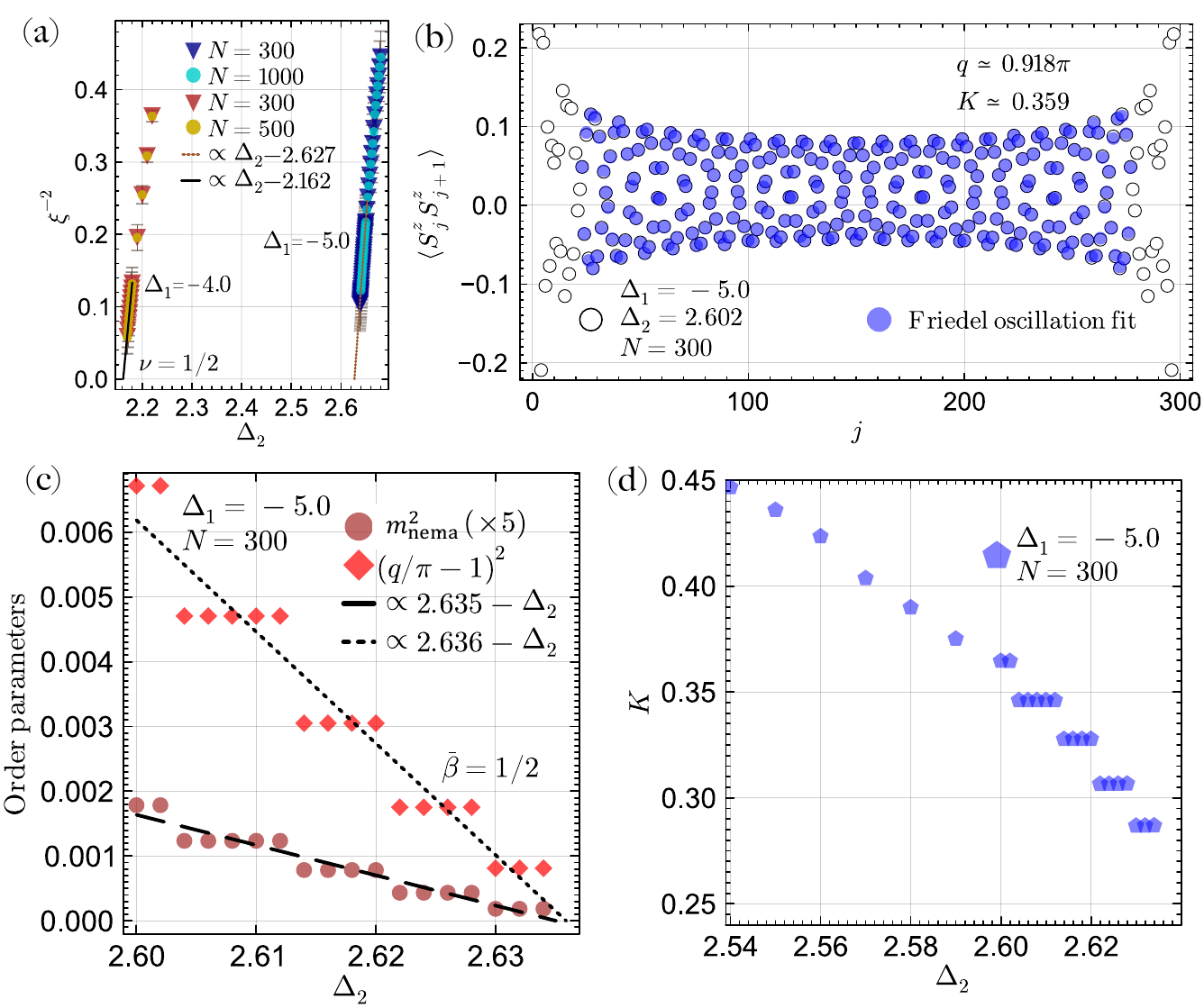} 
    \caption{Numerical evidences of Pokrovsky-Talapov transition from $c=2$ liquid phase to period-4 phase. (a) Scaling of the correlation length $\xi$ extracted from $\langle S^{z}_{i}S^{z}_{i+1}S^{z}_{j}S^{z}_{j+1}\rangle$ correlations for $\Delta_1=-4.0$, $\Delta_1=-5.0$ and several system sizes in the period-4 gapped phase according to Eq.~\eqref{OZ}. The results are in excellent agreement with theory prediction $\nu=1/2$ (black and brown lines). (b) Friedel oscillations in the $c=2$ liquid induced by boundary conditions $S^z_1 = -1/2$, $S^z_N = 1/2$. The fit to Eq.~\eqref{FO} yields values for the wave vector $q\simeq0.918\pi$ and the Luttinger parameter $K\simeq0.359$. (c) Scaling of the deviation of the wave vector from commensurate value $\Delta q=|\pi-q |$ (diamonds) and of the nematic density $m_{\mathrm{nema}}$ (circles) in the $c=2$ phase. The finite step value of the nematic density $\Delta m_{\rm{nema}}\simeq 1/N$ is the main text. The fit shows good agreement with the critical exponent $\bar{\beta}=1/2$. (d) Luttinger parameter $K$ approaching the critical value $K_c=1/4$ at the transition point $(\Delta_1=-5.0,\Delta_2\simeq2.63)$.}
    \label{RPT_tot}
	\vspace{-10pt}
\end{figure}

\paragraph{Pokrovsky-Talapov transition: from a gapped period-4 phase to a $c=2$ liquid.}
This transition continuously closes the gap of the period-4 $\downarrow\downarrow\uparrow\uparrow\dots\downarrow\downarrow\uparrow\uparrow$ phase (stabilized when $\Delta_2$ dominates: NNN spins anti-align along $z$), giving rise to a incommensurate critical phase with central charge $c=2$. As discussed later, the latter is interpreted as the coexistence of two LLs: one with standard single-magnon excitations and an incommensurate component made of kink-antikink pairs. Fig.~\ref{RPT_tot} summarizes the main features. 
In the gapped phase, we extract the correlation length $\xi$ by fitting a correlation function $G(x,0)$, e.g. $\langle S^+(x)S^-(0)\rangle$, to an Ornstein-Zernike form~\cite{OZ_ref}
\begin{equation}  
\label{OZ}  
G(x,0)\propto \frac{e^{-x/\xi}}{x^{1/2}}.  
\end{equation}  
The extracted correlation length allows us to locate the transition. The critical scaling \eqref{ξ_scaling} is presented in Fig.~\ref{RPT_tot}(a) and is in excellent agreement with Pokrovsky-Talapov critical exponent~\cite{PT} $\nu=1/2$:
\begin{equation}  
\label{ξ_scaling}
\xi\propto | g-g_{c} |^{-\nu},  
\end{equation}
where $g$ is a parameter which drives the transition (e.g. $\Delta_1$). In the gapless phase, we fit  Friedel oscillations in a density profile $n(x)$ for chain length $N$ with polarized boundaries with~\cite{hikihara2001}
\begin{equation}
\label{FO}
n(x) \propto \frac{\cos{\left(q x\right)}}{\left(\frac{2N}{\pi}\sin{\left(\frac{\pi x}{N}\right)}\right)^K}+\mathrm{cst}.
\end{equation}
From such fits (see Fig.~\ref{RPT_tot}(b)), we extract the wave vector $q$ and the LL coefficient $K$. Fig.~\ref{RPT_tot}(c) shows that close to the transition, the deviation $\Delta q = |q - q_C|$ from the commensurate wave vector $q_C$ scales with a critical exponent $\bar{\beta} = 1/2$ according to
\begin{equation}
\label{Δq_scaling}
\Delta q \propto | g-g_{c} |^{\bar{\beta}}.
\end{equation}
We identify the nematic density $m_{\mathrm{nema}}=\sum_i\langle S^z_iS^z_{i+1}\rangle/N$ as the condensed quantity (vanishing in the period-4 phase) responsible for the incommensurability of the critical phase. Thus, we also present the scaling $m_{\mathrm{nema}} \propto | g-g_{c} |^{\bar{\beta}}$, coming from the relation $\Delta q=2\pi m_{\mathrm{nema}}$~\cite{SM} between wavevector and density. Finally, Fig.~\ref{RPT_tot}(d) shows the value of $K$ approaching the field theoretical critical value $K_c=1/4$~\cite{giam}.

In the End Matter, we perform a similar numerical analysis of the second PT transition, between the gapped phase separation and nematic LL. As shown in the End Matter, lifting the $m=0$ constraint maps the gapped phase separation onto a ferromagnetic (FM) phase and the nematic LL condenses in the same manner, leaving the PT critical line unchanged ~\cite{SM}. 

\begin{figure}
    \centering
    \includegraphics[width=\columnwidth]{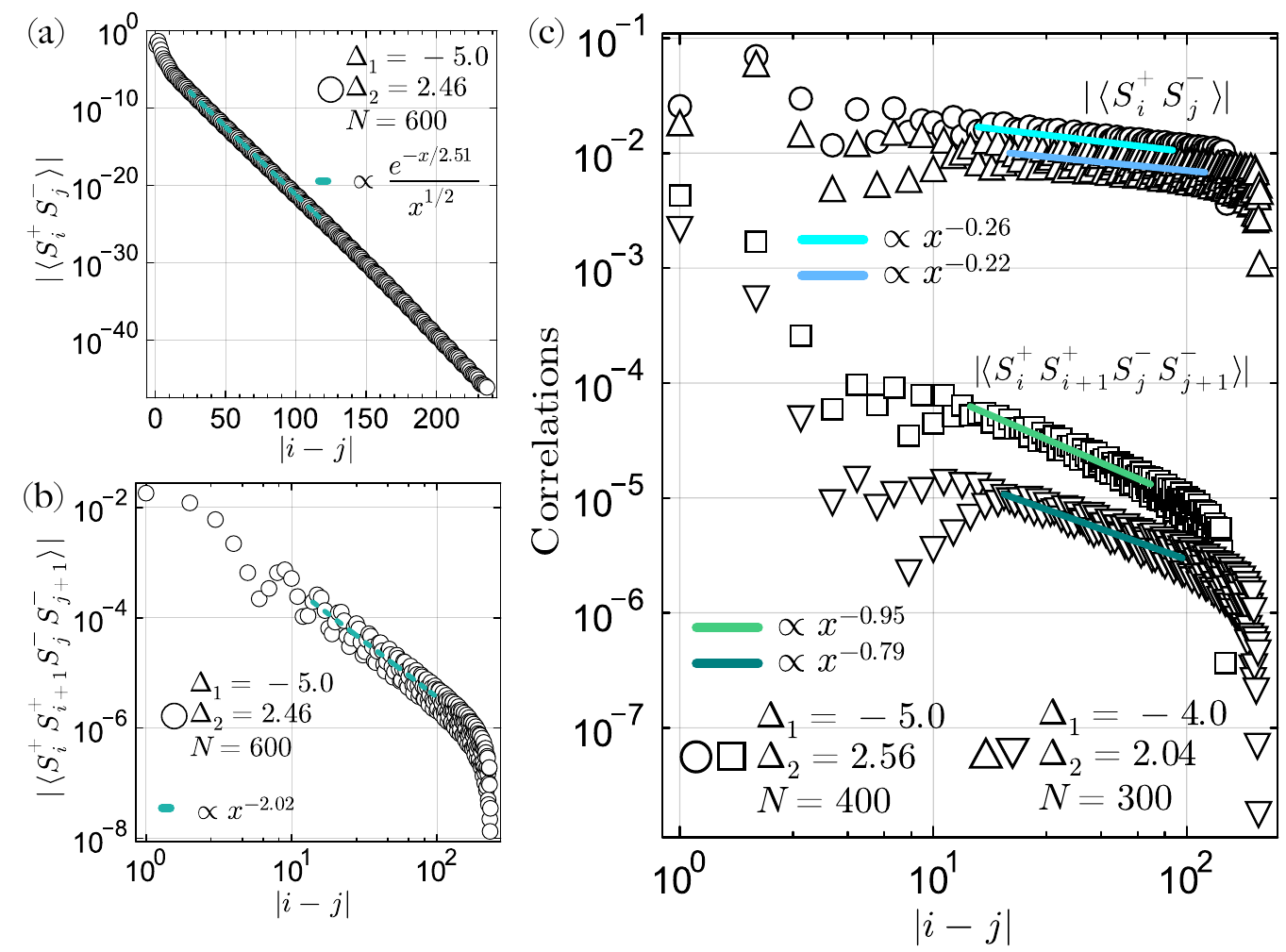}
    \caption{Spin flip correlations. (a) Single-spin flip correlations in the nematic Luttinger liquid phase. The exponential decay (see Eq.~\eqref{OZ}) reveals a gap in the excitation spectrum for single-spin flip events, as breaking a bound state of two spins costs a finite energy. (b) Two-spin flip (or nematic) correlations at the same parameter point. These correlations exhibit algebraic decay, reflecting the gapless nature of the two-bound magnon Luttinger liquid. (c) Single-spin and two-spin flip correlations (at two different parameters point) in the $c=2$ phase; both are of the algebraic type.}
    \label{SpSm_corrs}
\end{figure}

\begin{figure}
    \centering
    \includegraphics[width=\columnwidth]{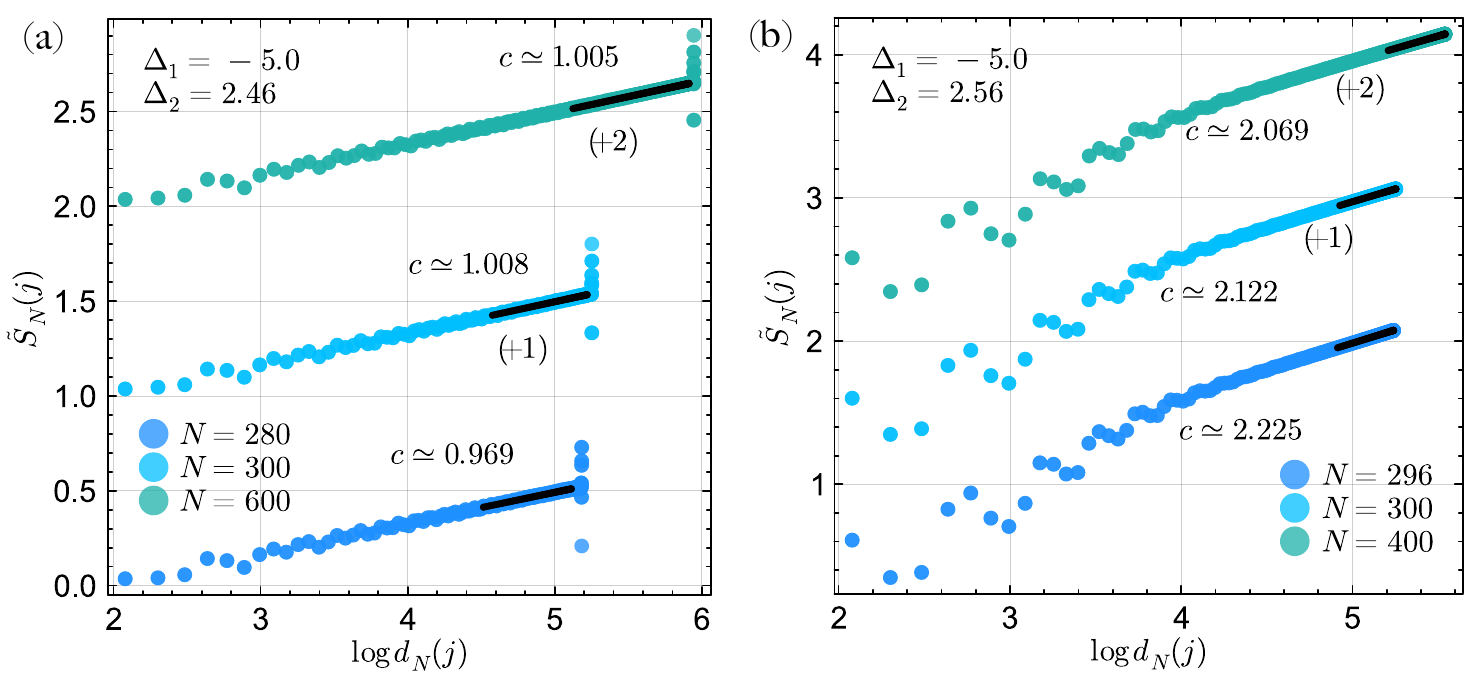}
    \caption{Scaling of the reduced entanglement entropy $\tilde{S}_N(j)$ as a function of the logarithm of the conformal distance $d_N(j)=(2N/\pi)\sin{\left(\pi x/N\right)}$ according to Eq.~\eqref{CC}. As the system size increases, the central charge approaches the value (a) $c=1$ in the nematic LL and (b) $c = 2$ in the second critical phase. The very last points for each data set in (a) correspond to $\tilde{S}_N(j)$ values near the middle of the chain where phase separation occurs (see End Matter) and are therefore excluded from the fit.}
    \label{c_LLs}
    \vspace{-10pt}
\end{figure}

\paragraph{Nematic Luttinger liquid vs $c=2$ phase.}
A major signature of the nematic LL is the short-range character of the correlation function $\langle S^+_iS^-_j\rangle$, while the correlations $\langle S^+_{i}S^+_{i+1}S^-_{j}S^-_{j+1}\rangle$ exhibit quasi-long-range order (see Fig.~\ref{SpSm_corrs}(a)-(b)). This indicates a finite energy gap for exciting the system via a single-spin flip, and gapless excitations for two-spin flip events. In contrast, both correlation functions are critical in the second phase, as shown in Fig.~\ref{SpSm_corrs}(c). The two gapless phases also differ by their central charge. This is extracted from the scaling of the entanglement entropy $S_N(x)$ along a chain of length $N$ using the Calabrese-Cardy formula~\cite{CC_ref}:
 \begin{equation}
 \label{CC}
 S_N(x)=\frac{c}{6}\log\left(\frac{2N}{\pi}\sin\left(\frac{\pi x}{N}\right)\right)+\log{g}+C,
 \end{equation}
 where $\log g$ stands for boundary entropy and $C$ is a non-universal constant. To extract the central charge from a linear fit, we first remove the Friedel oscillations contributing to $S_N(x)$ by defining $\tilde{S}_N(x) = S_N(x) + an(x)$, where $n(x)$ is a density profile and $a$ is a tuning parameter~\cite{laflorencie,capponi}. Fig.~\ref{c_LLs} shows that the incommensurate nematic phase has $c=1$, while the other liquid has $c=2$. The structure factor analysis in the End Matter further suggests that this $c=2$ phase hosts both single-magnon and incommensurate gapless modes.

Another distinctive feature of the nematic LL phase at zero magnetization is its phase separation profile. However, we show in the End Matter that phase separation is not a prerequisite for stabilizing the nematic LL phase, which emerges even when the magnetization constraint is lifted. In such a case, it appears in direct contact with the fully ferromagnetic phase, similarly to what is observed in~\cite{multipolar_LL,lauchli,sakai}.

\paragraph{Discussion.}
To gain an intuitive understanding of the origin of these PT transitions, we present a simple analytical description that allows us to predict the location of both PT lines. We work in the strongly interacting regime $|\Delta_1|,\Delta_2\gg J$ so that we write the Hamiltonian~\eqref{H_spin} as $H=H_0+JV$, where $H_0=\sum_j\big(\Delta_1 S^z_jS^z_{j+1}+\Delta_2 S^z_jS^z_{j+2}\big)$ and $V=\frac12\sum_j(S^{+}_jS^{-}_{j+1}+{\rm H.c.})$. We focus near the classical line $\delta := \Delta_2+\Delta_1/2 = 0$, which in the classical limit $J=0$, separates the two classical phases -- FM and period-4. The classical degeneracy at $\delta=0$ is lifted by quantum fluctuations at $J>0$~\cite{multiphase}. For the PT transition between the FM gapped phase and the nematic LL, we assume slightly negative detuning $\delta<0$ and lowest excitations out of the FM vacuum as bound two-magnon (quadrupolar) objects. Using degenerate perturbation theory~\cite{cohen} (equivalently, a second-order Schrieffer-Wolff reduction~\cite{SW,SW_2}), we block diagonalize $H$ with respect to the low-energy subspace spanned by hard-core bosons $b^\dagger_j =  S^{-}_{j}S^{-}_{j+1}$ on bonds, subject to the constraint $n_jn_{j+1}=0$, where $n_j=b^\dagger_j b_j$~\cite{competing_sachdev}. This low-energy subspace restriction is controlled because the intermediate broken-pair states lie higher by an $O(\Delta_2-\Delta_1)$ gap. Integrating out the space of virtual broken-pair states to $O(J^2)$, we obtain the effective bond lattice bosonic Hamiltonian~\cite{SM}
\begin{equation}
H_{\rm eff}= \sum_j \varepsilon \, n_j-t \, (b^\dagger_{j}b_{j+1}+{\rm H.c.}) +\sum_{r=2}^{4}V_r \, n_j n_{j+r},
\label{eq:Heff_PRL}
\end{equation}
where $-\varepsilon=2\delta+J^2/2(\Delta_2-\Delta_1)$ is an effective chemical potential for two-magnon pairs, $t=J^2/4(\Delta_2-\Delta_1)$ is the effective pair hopping and $V_{2,3,4}$ are interaction amplitudes. At infinitesimal pair filling relevant to the PT onset, the particles are non-interacting, hence the interaction terms $V_{r}$ do not affect the one-particle gap~\cite{giam}. The critical point is a vacuum instability of two-magnon bound pairs of density $\rho_{b}=\sum_j \langle n_j\rangle /N=0$ on the FM side. It occurs when the bottom of the dispersion band $E(k)=\varepsilon-2t\cos (k)$ closes~\cite{giam,PT}, i.e. $\varepsilon=2t$. We obtain the PT line equation
\begin{equation}
\Delta_{2,c}=\frac{\Delta_1}{4}\left[1-3\left(1-\frac{8J^2}{9\Delta_1^2}\right)^{\frac12}\right] < -\frac{\Delta_1}{2}.
\label{LPT_line}
\end{equation}
The line appears in pink in Fig.~\ref{PD_OH_v7} and is in excellent agreement with the critical points determined numerically. Physically, each two-magnon bound pair corresponds to a two-spin flip out of the fully polarized FM state. The magnetization density reads $m=1/2-2\rho_b$ and total magnetization changes by successive $\Delta M=2$ steps (as shown in the End Matter). Moreover, incommensurate modulation $q_{\rm{IC}}$ of local spin $S^z_j$ arises from a finite bound-pair density. In the dilute regime, each pair changes the nematic density by $\Delta m_{\rm{nema}}=-1/N$, hence $m_{\rm{nema}}=1/4-\rho_b+O(\rho_b^2)$. It follows the relation $q_{\rm{IC}}\simeq2\pi (1/4-m_{\rm{nema}})$, which is verified numerically~\cite{SM}.

In the End Matter, we show that a similar perturbation analysis in the $\delta > 0$ sector explains the PT transition out of the gapped period-4 phase in terms of proliferation of discommensuration kink-antikink pairs which interpolate between the spin patterns $\downarrow\downarrow\uparrow\uparrow\cdots$ and $\uparrow\uparrow\downarrow\downarrow\cdots$. The predicted PT line reads
\begin{equation}
\Delta_{2,c}=\frac{\Delta_1}{4}\left[1-3\left(1+\frac{16J^2}{9\Delta_1^2}\right)^{\frac12}\right] >-\frac{\Delta_1}{2}.
\label{RPT_line}
\end{equation}
This line is shown in green in the phase diagram in Fig.~\ref{PD_OH_v7} and is in very good agreement with the critical points determined numerically. The mechanism outlined above is also consistent with the other numerical results. For a finite chain with pinned ends $S^z_1=-S^z_N=1/2$, solitons enter as kink-antikink pairs, removing two domain walls. Writing the local nematic operator as $S^z_jS^z_{j+1}=(1-2n_j)/4$ where $n_j \in \{0,1\}$ is the local domain wall operator, this corresponds to a nematic density step $\Delta m_{\rm{nema}} = 1/N$, as shown in Fig.~\ref{RPT_tot}. Moreover, a kink (antikink) carries a magnetization  $\Delta M=+2$ $(-2)$; as kink-antikink solitons are condensed pairwise, the magnetization density remains $m=0$. As shown numerically~\cite{SM}, this holds even without magnetization constraint. Finally, because a kink interpolates between the two period-4 patterns, the latter differ by a translation by two sites and thus by a phase shift $\Delta \phi=\phi(\infty)-\phi(-\infty)=\pi$. Writing the local magnetization as $S^z(x)\sim \cos(q_0x+\phi(x))$ where $q_0=\pi/2$ (period-4 pattern), the kink density is $\rho_s=\partial_x\phi/\pi$ and we obtain $|q_{\rm{IC}}-\pi/2|=\pi \rho_s$. In the dilute-kink regime relevant to the PT onset, the domain wall occupation reads $\bar{n}=1/2-\rho_s+O(\rho_s^2)$, which for the nematic density gives $m_{\rm{nema}}=\rho_s/2+O(\rho_s^2)$. We thus obtain the relation $|q_{\rm{IC}}-\pi/2|\simeq 2\pi m_{\rm{nema}}$, which is verified numerically~\cite{SM}.

\paragraph{Conclusion.} We report a very rich phase diagram of a quantum XXZ spin-1/2 chain with nearest-neighbor ferromagnetic exchange $\Delta_1$ and next-nearest-neighbor antiferromagnetic exchange $\Delta_2$ at fixed zero magnetization and without magnetization constraint. We have discovered two Pokrovsky-Talapov commensurate-incommensurate transitions in the strongly interacting regime, leading respectively to the condensation of an incommensurate phase with central charge $c=2$ and an incommensurate nematic Luttinger liquid. We interpret the former as comprising both a single-magnon gapless mode and an incommensurate one. The most surprising aspect is that these two transitions occur in the absence of any magnetic field or doping as opposed to a generically accepted theory~\cite{giam}. Instead, these Pokrovsky-Talapov transitions are enabled by the competition between $\Delta_1$ and $\Delta_2$ when both are strong enough. We also provide numerical and analytical evidence for the nature of these gapless phases: the first arises from the continuous softening and subsequent proliferation of kink-antikink discommensuration pairs out of an antiferromagnetic-like gapped phase, whereas the second marks the condensation of two-magnon bound pairs as the ferromagnetic-like gap closes continuously. This pairwise condensation of quasiparticles, or more generally, the condensation of complex objects, defines an alternative mechanism for commensurate-incommensurate Mott transitions. The associated incommensurate critical phases arise purely due to frustration, without any external magnetic field -- an exciting scenario that, to our knowledge, has not been reported before.

\vspace{15pt}
\paragraph{Acknowledgements}
J. F. is indebted to Bowy La Rivière for providing his DMRG code used in this work, and thanks Edmond Orignac for valuable discussions. N. C. acknowledges useful discussions with Thierry Giamarchi and Philippe Lecheminant. This research has been supported by the Julian Schwinger Foundation. N. C. acknowledges support from the Royal Society (grant number URFR1251326). Numerical simulations have been performed with the Dutch national e-infrastructure with the support of the SURF Cooperative and at the DelftBlue HPC.

\newpage

\section{End Matter}

\begin{figure}[h]
    \centering
    \includegraphics[width=\columnwidth]{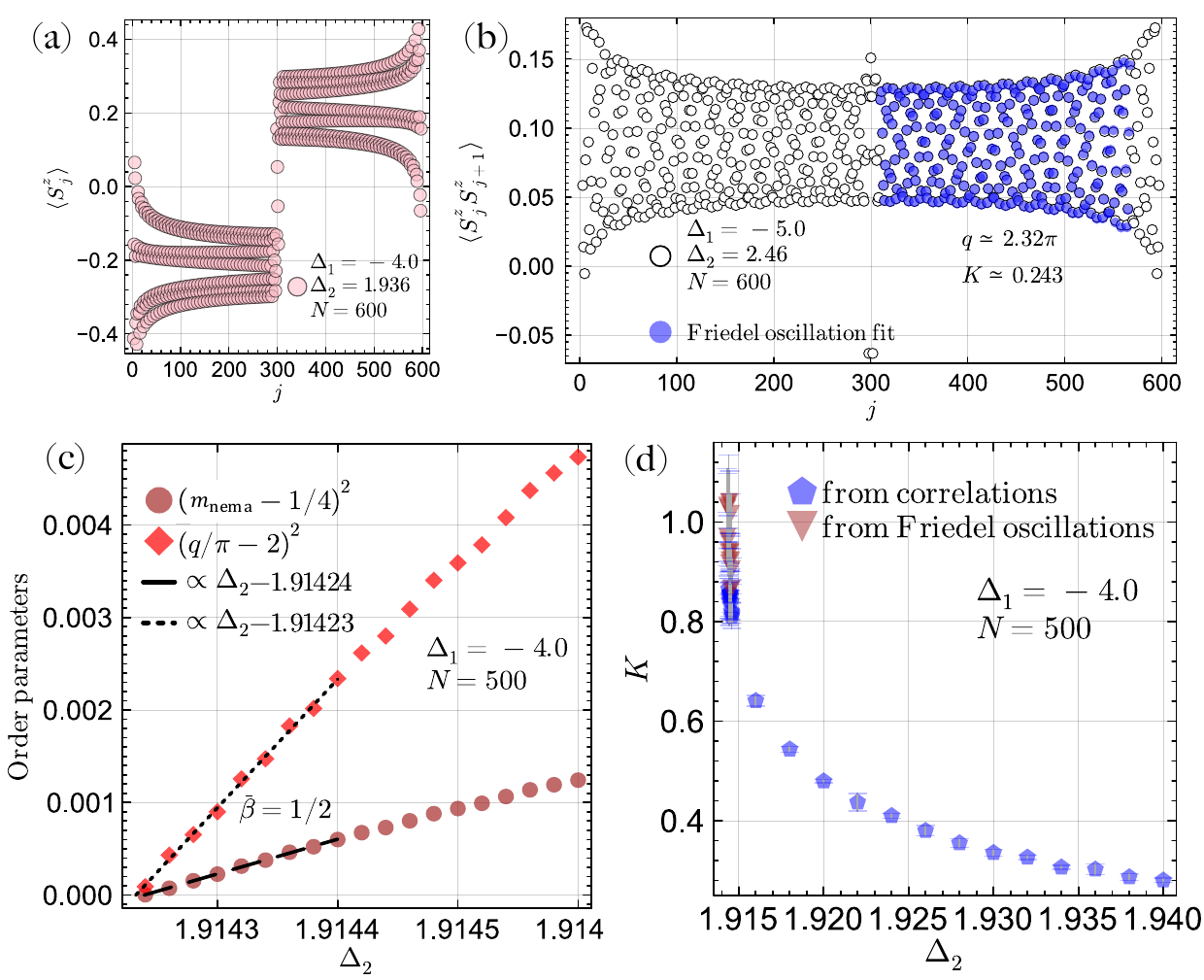}
     \vspace{-15pt}
    \caption{Numerical evidences of the Pokrovsky-Talapov transition between the gapped phase separation and nematic Luttinger liquid (LL). (a) Incommensurate phase separation profile $\langle S^z_j\rangle$ in the nematic LL. (b) Friedel oscillations in the nematic LL induced by open boundary conditions. The fit with Eq.~\eqref{FO} in the main text yields values for the wave vector $q\simeq2.32\pi$ and the Luttinger parameter $K\simeq0.243$. (c) Scaling of the wave vector deviation $\Delta q=|2\pi-q |$ and of the nematic density $m_{\mathrm{nema}}$. The fit shows excellent agreement with the exponent $\bar{\beta}=1/2$. (d) Luttinger parameter $K$ approaching the critical value $K_c=1$ at the transition point ($\Delta_1=-4.0,\Delta_2\simeq1.914$).}
    \label{LPT_tot}
\end{figure}

\begin{figure}
    \centering
    \includegraphics[width=\columnwidth]{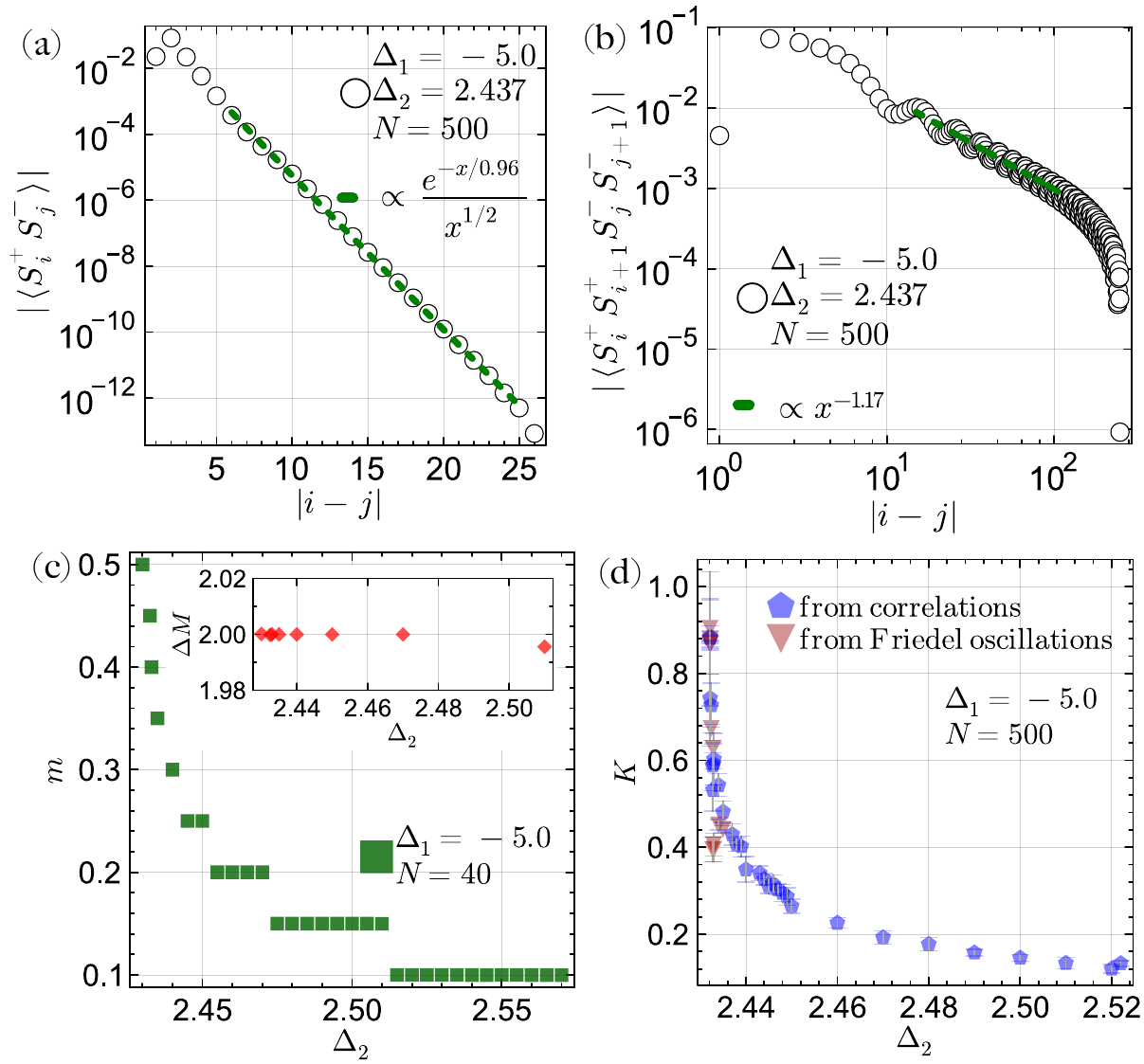}
    \vspace{-5pt}
    \caption{Properties of the nematic Luttinger liquid in the system with free magnetization. (a)-(b) Single-spin-flip versus two-spin flip (nematic) correlations: single-spin flip processes are gapped, while two-spin flips are gapless. (c) Magnetization near the Pokrovsky-Talapov transition. The first point corresponds to the ferromagnetic gapped state, $m = 1/2$. The step size $\Delta M = N \Delta m = 2$ indicates that $m$ changes via two-spin flips. (d) Luttinger parameter $K$, calculated from correlations~\eqref{corr_boso} and Friedel oscillations induced by $S^z_1=S^z_N=1/2$, approaching the critical value $K_c = 1$ at the transition point.}
    \label{LPT_NLL_FREE}
\end{figure}

\paragraph{\hspace{-10pt}Pokrovsky-Talapov transition: from phase separation to nematic Luttinger liquid.}
Fig.~\ref{LPT_tot} summarizes the main features. As for the previous PT transition between gapped period-4 and $c=2$ phases, we show the critical scaling satisfying $\bar{\beta}=1/2$ for $\Delta q$ and for $m_{\mathrm{nema}}$. Fig.~\ref{LPT_tot}(d) shows the value of $K$ extracted from Friedel oscillations (Fig.~\ref{LPT_tot}(b)) approaching the field theoretical critical value $K_c=1$~\cite{giam}. $K$ is also measured from the nematic correlations $\langle S^{+}_iS^{+}_{i+1}S^{-}_{j}S^{-}_{j+1}\rangle$ upon fitting with the first term of the field theory prediction~\cite{multipolar_LL}
\begin{equation}
\label{corr_boso}
\langle S^{+}_iS^{+}_{i+1}S^{-}_{j}S^{-}_{j+1}\rangle= \frac{A}{|i-j|^{1/2K}}+\frac{B\cos{\left(2Q|i-j|\right)}}{|i-j|^{2K+1/2K}},
\end{equation}
where $A$ and $B$ are non-universal constants and $Q$ is a wave vector related to the two-magnon bound pair filling. An example of such fit is shown in Fig.~\ref{LPT_NLL_FREE}(b).

\paragraph{Releasing the magnetization constraint~\cite{itensor,itensor-r0.3}\footnote{Numerical simulations reported in this and the following paragraphs have been performed using the ITensor Julia library.}.}
The only phases that are affected by the constraint are the gapped phase separation (which becomes fully ferromagnetic) and the nematic LL. All other phases still have $m=0$ when the magnetization is free. In these former phases, the only effect of the constraint is to flip the sign of all spins in one half of the chain and to introduce a localized domain wall at the center of the chain. The healing length that defines the width of this wall, $w_{\mathrm{PS}}$, becomes of the order of the lattice spacing for sufficiently large system sizes ($N \gtrsim 100$), so that the thermodynamic limit behavior $\lim_{N \to \infty} w_{\mathrm{PS}}(N)/N = 0$ is rapidly reached.
 As shown in Fig.~\ref{LPT_tot}(a)-(b), it is already negligible ($\simeq 4$ sites) for $N=600$. Thus, the constraint has no impact on the physics of interest, namely the key properties relevant to describing the phase transitions. In particular, the local nematic density $\langle S^z_{i}S^z_{i+1}\rangle$ in the nematic LL is sensitive only to the relative sign of neighboring spins and is therefore unaffected, up to the negligible influence of the domain wall.
Fig.~\ref{LPT_NLL_FREE}(a)-(b) shows the nematic features of the nematic LL correlations in the unconstrained system. Figure~\ref{LPT_NLL_FREE}(c) shows the behavior of the magnetization $m$ near the transition, whose gradual deviation from $m=1/2$ causes the incommensurability of the phase. This variation occurs through successive two-spin flips typical of the nematic phase, as indicated by the step size $\Delta M = 2$ in the total magnetization $M = Nm$. It involves nematic condensation via $\Delta m_{\mathrm{nema}}=1/N$ steps. The PT nature is further supported by the Luttinger parameter calculation approaching $K \rightarrow K_c = 1$ at the transition, as shown in Fig.~\ref{LPT_NLL_FREE}(d). Nematic density PT scaling at several $N$ is presented in the Supplemental Material~\cite{SM} and shows that the absence of constraint plays no role in the transition location.
\begin{figure}
    \centering
    \includegraphics[width=\columnwidth]{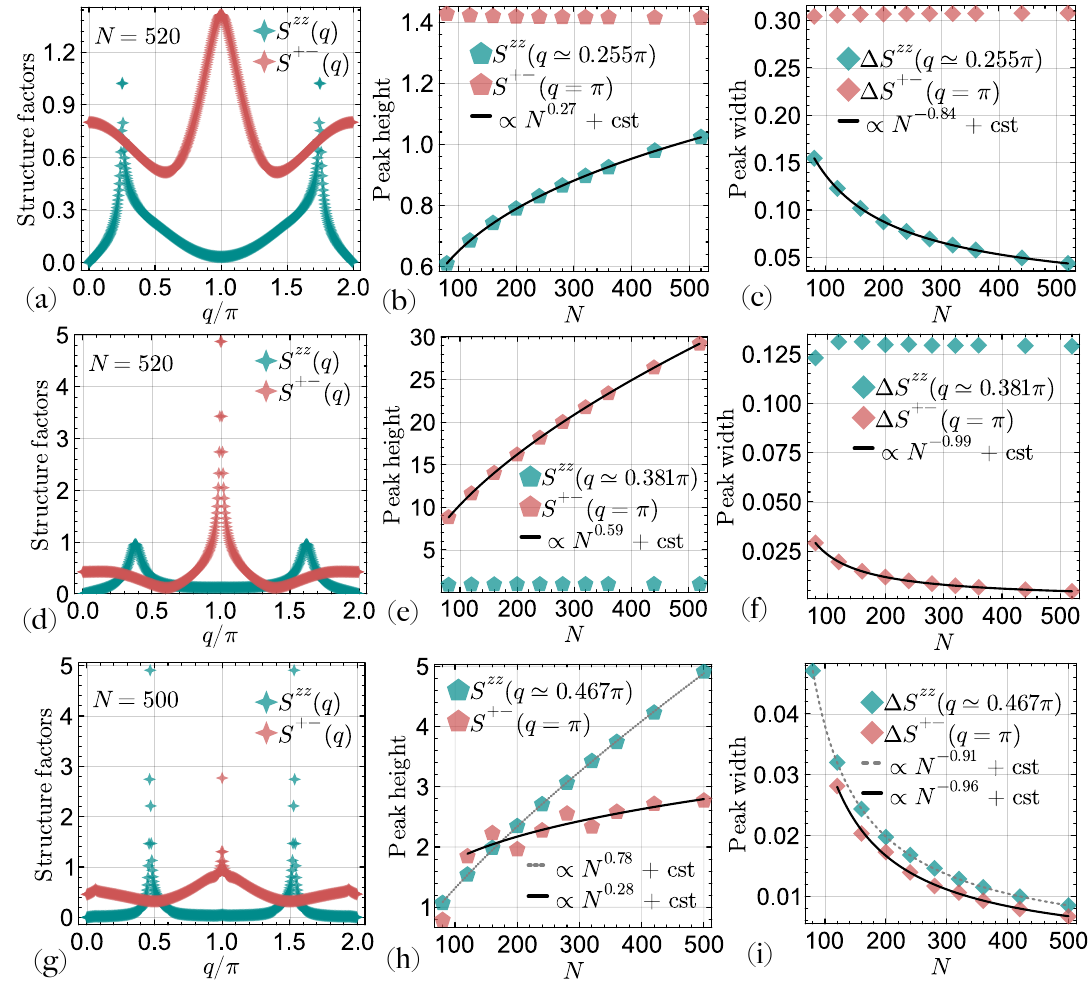}
       \vspace{-20pt}
    \caption{Structure factor features of the three critical phases. Panels~(a)-(c) show, for the nematic Luttinger liquid at $(\Delta_1,\Delta_2)=(-2.68,1.2272)$, (a) $S^{zz}$ and $S^{+-}$, (b) the finite-size scaling of the peak heights, and (c) the finite-size scaling of the peak widths. Panels~(d)-(f) show the same quantities for the single-magnon-type Luttinger liquid at $(\Delta_1,\Delta_2)=(-2.0,1.0)$, and panels~(g)-(i) for the $c=2$ phase at $(\Delta_1,\Delta_2)=(-2.68,1.54)$.}
           \vspace{-20pt}
    \label{Sq}
\end{figure}

\paragraph{Criticality of the $c=2$ phase.}
To analyze the gapless modes in the $c=2$ phase, we compute the structure factors $S^{zz}(q)$ and $S^{+-}(q)$ of the three critical phases in the longitudinal ($zz$) and transverse ($+-$) channels, respectively. We use the definition~\cite{SQ_Sandvik,SQ_Kumar}
\begin{equation}
S^{\alpha\beta}(q)=C^{\alpha\beta}(0)+2\sum_{r=1}^{N-1}\cos(qr)\;\mathrm{Re}\;C^{\alpha\beta}(r),
\end{equation}
where $C^{\alpha\beta}(r)=\frac{1}{N-r}\sum_{j}C^{\alpha\beta}_{j,j+r}$ is the pair-averaged correlator at separation $r=|i-j|$, $C^{\alpha\beta}_{ij}=\langle S^{\alpha}_iS^{\beta}_j\rangle-\langle S^{\alpha}_i\rangle\langle S^{\beta}_j\rangle$ is a connected correlation function and $\alpha,\beta=z,+$, or $-$. Fig.~\ref{Sq} summarizes the structure factor signatures of the three critical phases. The incommensurate peak in the density ($zz$) channel reveals the only gapless mode of the nematic LL, whose height diverges as the expected power law with system size $N$~\cite{SQ_Sandvik,SQ_Karbach,SQ_Suman}, while its width decreases as $\sim N^{-1}$. In contrast, the $q=\pi$ peak of the transverse ($+-$) structure factor shows gapped excitation: both its height and width quickly saturate to constant values (once $N$ is sufficiently larger than the correlation length~\cite{SQ_Kumar}). The standard LL exhibits the opposite features: the $q=\pi$ peak of $S^{+-}$ captures the only gapless mode available in a single-magnon-type LL, namely the single-spin-flip channel, while the incommensurate excitation revealed in $S^{zz}$ is gapped. By comparison, in the $c=2$ phase, both the incommensurate peak and the $q=\pi$ single-spin-flip peak sharpen with system size by exhibiting a power-law growth of their heights and a $\sim N^{-1}$ narrowing of their widths. This simultaneous critical finite-size scaling of both peaks is suggestive of the presence of two critical modes.

\paragraph{Pokrovsky-Talapov from gapped period-4 phase to $c=2$ liquid: perturbation analysis.}
On the period-4 side $\delta = \Delta_2+\Delta_1/2>0$ of the classical line $\delta = 0$, we first introduce bond-center domain-wall occupations $n_{j}=(1-4S^z_jS^z_{j+1})/2\in\{0,1\}$, for which the unperturbed Hamiltonian reads $H_0=-\delta\sum_j n_j+\Delta_2\sum_j n_j n_{j+1}$. Since $\Delta_2>0$ penalizes adjacent domain walls, the low-energy sector is restricted by the hard constraint $n_j n_{j+1}=0$. For $\delta>0$, $H_0$ is minimized by the two wall patterns $0101\cdots$ and $1010\cdots$ with a domain-wall occupation $\bar{n}=\sum_j\langle n_j\rangle/N=1/2$. Solitons are discommensurations (kinks) that interpolate between the spin patterns $\downarrow\downarrow\uparrow\uparrow\cdots$ and $\uparrow\uparrow\downarrow\downarrow\cdots$, each of them removing one domain-wall. We denote by $b^\dagger_j$ the operator creating a kink on the bond-center site $j+1/2$ (a nonlocal operator in the original spins) and by $\nu_j= b^\dagger_j b_j$ its hard-core occupation. In the dilute-kink sector relevant at the PT threshold, the effective Hamiltonian to $O(J^2)$ reduces to a hard-core soliton gas on the bond lattice~\cite{SM}
\begin{equation}
H_{\rm eff}=\sum_j \varepsilon \, \nu_j
-t \, \big(b^\dagger_{j}b_{j+1}+\mathrm{H.c.}\big)
+V_1 \nu_j \nu_{j+1},
\end{equation}
where $-\varepsilon=-\delta+J^2/2(\Delta_2-\Delta_1)$ is the effective chemical potential for kinks and $t=J^2/4(\Delta_2-\Delta_1)$ is the hopping amplitude. The interaction $V_1$ is short-ranged and does not affect the location of the vacuum instability~\cite{giam}. The critical point corresponds to the closure condition $\varepsilon=2t$ of the dispersion band $E(k)=\varepsilon-2t\cos(k)$ bottom. This provides the equation line \eqref{RPT_line} mentioned in the main text.

\FloatBarrier

\bibliographystyle{apsrev4-2}
\bibliography{references}

\end{document}